\documentstyle[12pt]{article}

\def\be{\begin{equation}}
\def\ee{\end{equation}}
\def\ba{\begin{array}{c}}
\def\ea{\end{array}}

\def\ben{$$}
\def\een{$$}
\begin{document}

\titlepage
\vspace*{4cm}

 \begin{center}
{\Large \bf Conservation of pseudo-norm
\\
 in ${\cal PT}$ symmetric
quantum mechanics
 }

\end{center}

\vspace{5mm}

 \begin{center}

Miloslav Znojil
\vspace{3mm}

\'{U}stav jadern\'e fyziky AV \v{C}R, 250 68 \v{R}e\v{z},
Czech Republic\\

e-mail: znojil@ujf.cas.cz

\end{center}

\vspace{5mm}

\section*{Abstract}

We show that the evolution of the wave functions in the ${\cal
PT}$ symmetric quantum mechanics is pseudo-unitary. Their
pseudo-norm $\langle \psi| {\cal P} | \psi \rangle$ remains
time-independent. This persists even if the ${\cal PT}$ symmetry
itself becomes spontaneously broken.

\vspace{9mm}

\noindent
\noindent PACS 03.65.Bz, 03.65.Ca 03.65.Fd


 \begin{center}
 {\small \today, spon.tex file}
 \end{center}

\newpage

\section{Introduction}

Evolution in quantum mechanics \cite{Landau}
 \ben
 |\psi(t)\rangle
= e^{-iHt}\,
 |\psi(0)\rangle
 \een
conserves the norm of a state.  The assumption $H = H^\dagger$
of the Hermiticity of the Hamiltonian leads to the
time-independence of the probability density,
 \be
\langle \psi(t)| \psi(t) \rangle=
\langle \psi(0)|
e^{iH^\dagger\,t}\,e^{-iH\,t}
| \psi(0) \rangle=
\langle \psi(0)|  \psi(0) \rangle.
\label{dve}
 \ee
{\it Vice versa}, in the light of the Stone's theorem, the
unitarity of the evolution implies the Hermiticity of the
Hamiltonian.

Apparently, there is no space left for the non-Hermitian
Hamiltonians which were recently studied by Bender et al
\cite{BBjmp}-\cite{spont}. The latter formalism may prove
inspiring in field theory \cite{BMa} but, as an extended quantum
mechanics, it contradicts the Stone's theorem.  In what follows,
we intend to clarify this point.

Section 2 reviews a few basic features of the extended formalism
which replaces the Hermiticity $H = H^\dagger$ by its weakening $H
= H^\ddagger$.  The latter property (called ${\cal PT}$ symmetry)
is explained and several parallels between the symmetries
$H=H^\dagger$ and $H=H^\ddagger$ are mentioned.  Explicit ${\cal
PT}$ symmetric square well solutions \cite{SQW} are recalled as an
illustration of the whole idea.

The introductory part of section 3 recollects the regularized
spiked harmonic oscillator bound states of ref. \cite{hopt} as a
solvable example of a ${\cal PT}$ symmetric system which is
defined on the whole real line. With its orthogonality and
completeness properties kept in mind, we arrive at the first
climax of our paper and formulate the appropriate modification of
the conservation law (\ref{dve}) in the case of a general
non-Hermitian system with the unbroken ${\cal PT}$ symmetry.

At the beginning of section 4 the construction of the harmonic
bound states is extended to the domain of couplings where the
${\cal PT}$ symmetry of the wave functions becomes spontaneously
broken.  We show how all the Hilbert-space-like concepts of
orthogonality and completeness of the states may be generalized
accordingly.  In particular, the ${\cal PT}$ symmetric norm (or
rather pseudo-norm) proves so robust that the new conservation
law of the type (\ref{dve}) remains valid even in the
spontaneously broken regime where the energies cease to be real.

\section{${\cal PT}$ symmetric quantum mechanics}

The concept of the extended, non-Hermitian quantum mechanics with
the requirement of ${\cal PT}$ symmetry of its Hamiltonians grew
from several sources.  The oldest root of its appeal is the
Rayleigh-Schr\"{o}dinger perturbation theory.  Within its
framework, Caliceti et al \cite{Caliceti} have discovered that a
low-lying part of the spectrum of the manifestly non-Hermitian
cubic anharmonic oscillator $H= p^2 + x^2 + g\,x^3$ is {\em real}
for the {\em purely imaginary} couplings $g$.  This establishes
many formal analogies with Hermitian oscillators
\cite{Alvarez,ixna}.

A different direction of analysis has been accepted by Buslaev
and Grecchi \cite{BG} who emphasized and employed some parallels
between the Hermiticity and ${\cal PT}$ symmetry during their
solution of an old puzzle of perturbative equivalence between
apparently non-equivalent quartic anharmonic oscillators
\cite{Seiler}.

A mathematical background of the non-unique choice of the
phenomenological Hamiltonians with real spectra has been pointed
out, in non-Hermitian setting, by several authors \cite{BT}.
Bender and Milton \cite{QED} emphasized the relevance of the
unique analytic continuation of boundary conditions for the
clarification and consequent explanation of the famous Dyson's
paradox in QED \cite{Dyson}.

In the cubic anharmonic models $H= p^2 + x^2 +i\,g\,x^3$ the
reality of the energies at the sufficiently small $g$
\cite{Caliceti,Pham} resembles the quartic case.  Bender and
Boettcher \cite{BB} attributed this connection to the
commutativity of the Hamiltonian with the product of the complex
conjugation ${\cal T}$ (which mimics the time reversal) and the
parity ${\cal P}$, $H = {\cal PT} H {\cal PT}= H^\ddagger$. An
acceptability of this conjecture has been supported by a few
partially \cite{QES} as well as completely \cite{exact} exactly
solvable models.

In the physics community, a steady growth of acceptance of the
${\cal PT}$ symmetric models can be attributed to their possible
phenomenological relevance.  The cubic $H= p^2 + i\,x^3$ has been
found relevant in statistical physics \cite{Bessis} and its
non-linear perturbations $H= p^2 + (i\,x^3)^{1+\delta}$ were
studied in field theory \cite{BMb}.  In all these models, a key
argument that they can prove useful in some applications has been
based on the reality of their spectrum. This argument is slightly
misleading as we shall see in what follows.

\subsection{Solvable illustration: Square well}

The possibility of a spontaneous breakdown of the reality of the
energies has been recently studied via a ${\cal PT}-$symmetric
quartic oscillator  $H= p^2 + i\,g\,x+x^4$ \cite{Simsek}. In this
model, one spots the sequence of ``critical" couplings $g_n$ such
that, step by step, the lowest real pair of the bound state
energies $E_{2n}$ and $E_{2n+1}$ becomes converted into a complex
conjugate doublet beyond~$g = g_n$. Such a pattern may prove
characteristic for a fairly broad class of non-Hermitian
interactions. For the sake of simplicity of the whole discussion,
let us pick up the Schr\"{o}dinger bound state problem on a finite
interval,
 \be
\left [
-\frac{d^2}{dx^2} +
V(x)\right ]\,\psi_n(x)
= E_n\,\psi_n(x),
\ \ \ \ \ \ \ \ \ \
 \psi_n(-1)
= \psi_n(1)=0, \label{SQWL}
 \ee
equipped with one of the most elementary ${\cal PT}$ symmetric
forces $V(x) =i\,T^2\,{\rm sign}\, x$.  Then, the ansatz
 \ben
\psi_n(x) = \left \{ \ba \sin \lambda (x+1),  \ \ \ \ \ \ x < 0,\\
C\,\sin \kappa (x-1),  \ \ \ \ \ \ x < 0 \ea \right .
 \een
with $E = k^2$ defines the solutions via the matching condition at
$x=0$. Using an abbreviation $\lambda = p-i\,q$ and rules
 \ben
\lambda^2=k^2-i\,T^2,
\ \ \ \ \ \ \ \
\kappa^2=k^2+i\,T^2= \lambda^*
\een
we get the elementary matching condition
 \ben
\frac{\tan \lambda}{\lambda} = {\rm purely\ imaginary}
 \een
which is equivalent to the elementary rule
 \ben
q\, \sinh 2q= -p\,\sin 2p .
 \een
Its numerical solution has been discussed elsewhere \cite{SQW}.
Still, without any detailed numerical computations one can fairly
easily see that in the two-dimensional $p - q$ plane, the
left-hand-side function forms a valley with zero minimum along the
line $q=0$.  The right-hand-side periodically oscillates with an
increasing amplitude.  One can conclude that the positive
solutions $p(q)>0$ of the latter equation form an infinite family
of ovals which are symmetric with respect to the $p-$axis.  The
$n-th$ oval is confined between the zeros of the sine function,
i.e., between the two lines $p=(2n+1)\pi/2$ and $p=(2n+2)\pi/2$.
With the growth of $n = 0, 1, \ldots$, the ovals are longer as
their ends move farther and farther from the $p-$axis.  In the $n
\gg 1$ asymptotic region, we get the estimate $p_{end} \approx
(n+3/4)\pi$ and $q_{end} \approx {\rm ln}\, n$.

As long as the definition of $p$ and $q$ implies that $p =
T^2/(2q)$ we get the second curve which is a plain hyperbola in
$p - q$ plane. The final solutions (i.e, intersections of this
hyperbola with all the ovals) move close to the standard square
well solutions in the quasi-Hermitian limit where $n \gg T^2$.

At the opposite extreme, the two lowest real energies determined
by the lowest oval cease to exist for the sufficiently strong
imaginary part of the force, i.e., for $ T^2
> T^2_{crit} \approx 2\,p_{end}\,q_{end}$.  In the light of the
previous estimates, the values of these critical points will grow
with the number $n$ of the oval in question. At $n=0$ one has
$T^2_{crit} \approx 4.48$ \cite{SQW}.

\section{Models with unbroken ${\cal PT}$ symmetry}

We can summarize that the elementary and exactly solvable ${\cal
PT}-$symmetric square well model has a spectrum $E_n$ which
remains real in a certain non-empty interval of couplings $T \in
(0,T_0)$.  At the boundary $T = T_0$ with certain exceptional
features \cite{Herbst}, the lowest energy doublet $E_{0}$ and
$E_{1}$ merges into a single state. The most immediate
fructification of this experience lies in the possibility of its
transfer to the ${\cal PT}$ symmetric potentials on a ``more
realistic" infinite interval of coordinates.

\subsection{${\cal PT}$ symmetric harmonic oscillators
\label{HOlab}}

A nontrivial example which is solvable on the full real line is
the ${\cal PT}$ symmetric harmonic oscillator described by the
differential Schr\"{o}dinger equation of ref. \cite{hopt},
 \be
\left [
-\frac{d^2}{dx^2} +
\frac{G}{(x-i\delta)^2}
-2i\delta\,x
+ x^2
\right ]\,\psi_n(x)
= E_n\,\psi_n(x),
\ \ \ \ \ \ \ \ \ \ \psi_n(x) \in L_2(-\infty,\infty).
\label{HO}
 \ee
This equation with $G > -1/4$ can be interpreted as a confluent
hypergeometric equation where an elementary change of the
coordinate $x=r+i\,\delta$ eliminates all the unusual imaginary
terms. At the same time, due to the analyticity of such a
transformation, one can simply keep $r$ on the real line with  a
small complex half-circle circumventing the singularity in the
origin ($r=0$). Without any loss of generality we can then work
with the complex general solution of our equation. It is available
in closed form,
 \ben
\psi(x) = C_{(+)}r^{1/2-\alpha}e^{r^2/2} \ _1F_1 \left [
\frac{1}{4}(2-E-2\alpha), 1-\alpha; r^2 \right ] + \een
 \ben
\ \ \ \ \ \ \ \ \ \ \ \ \ \ \ \ \ \ \ \ \ +
C_{(-)}r^{1/2+\alpha}e^{r^2/2} \ _1F_1 \left [
\frac{1}{4}(2-E+2\alpha), 1+\alpha; r^2 \right ].
 \een
This facilitates the use of the asymptotic boundary conditions. In
the standard way described in any textbook \cite{Fluegge} the idea
works without alterations since the general solutions grow
exponentially unless one of the confluent hypergeometric series
terminates to a Laguerre polynomial. This gives the compact wave
functions
 \ben
\psi_N(r)={\cal N}\,r^{1/2-Q\,\alpha}\,e^{-r^2/2}\,
L_n^{(-Q\,\alpha)}(r^2)
 \een
with the quasi-parity $Q = \pm 1$, main quantum number $n = 0, 1,
\ldots$ and subscripted index $N = 2n+(1-Q)/2$.  The energies
 \be
E_N=4n+2-2\,Q\,\alpha,
 \ \ \ \ \ \ \ \ \ \ \ \alpha = \sqrt{G+\frac{1}{4}}
 , \ \ \ \ \ \ \ \ G > -\frac{1}{4}
\label{spectrum}
 \ee
remain real for the positive centrifugal-like parameters
$\alpha>0$. These energies decrease/grow with $G$ for the
quasi-even/quasi-odd quasi-parity $Q$ of the state, respectively.

If necessary, we can return to the Hermitian case and reproduce
the usual radial harmonic oscillator solutions, provided only that
we cross out the ``unphysical" quasi-even states. These states
violate the textbook boundary conditions \cite{Landau}. The only
exception concerns the regular limit (one-dimensional case) with
$G=0$ (i.e., $\alpha=1/2$) and two parities $Q=\pm 1$. In this
sense, the full spectrum (\ref{spectrum}) of our complexified
oscillator represents a comfortable formal link between the
seemingly different one- and three-dimensional Hermitian
cases~\cite{creation}.

\subsection{Scalar product and orthogonality}

Let us return to a general Hamiltonian $H = H^\ddagger$ and
assume that its spatial parity ${\cal P}$ becomes manifestly
broken, $ {\cal P} H {\cal P} = {\cal T} H {\cal T} \neq H$.  We
may define the quasi-parity as a constant integer $Q_n= (-1)^n$
in the $n-$th state. This generalizes the above square well and
harmonic oscillator constructions and applies also to the
quartic oscillator of ref. \cite{Simsek} in a constrained domain
of the couplings $g \in  (0, g_0)$.

In the first step we introduce the indeterminate scalar product
defined by the prescription $\langle \phi | {\cal P} | \psi
\rangle$ of ref. \cite{whatis}.  It does not possess the property
of definiteness and defines merely a pseudo-norm. The
disappearance of the self-overlap $\langle \psi | {\cal P} | \psi
\rangle =0$ does not imply that the vector $ | \psi \rangle$ must
vanish by itself.

The main merit of such a definition of the scalar product lies in
the observation that it leads to the usual orthonormality of the
left and right eigenstates of the ${\cal PT}$ symmetric
Hamiltonians and Schr\"{o}dinger equations~\cite{ixna},
 \ben
\langle \psi_n | {\cal P}| \psi_m \rangle = Q_n \delta_{mn},
\ \ \ \ \ \ m , n = 0, 1, \ldots.
 \een
The completeness relations also acquire the form mentioned in ref.
\cite{whatis},
 \ben
\sum_{n=0}^\infty
| \psi_n \rangle \, Q_n
\langle \psi_n |
{\cal P} =  I.
 \een
The related innovated spectral representation of our
non-Hermitian ${\cal PT}$ symmetric Hamiltonians can be written
in a bit unusual but fully transparent manner,
 \ben
H=
\sum_{n=0}^\infty
| \psi_n \rangle \, E_n\,Q_n
\langle \psi_n |
{\cal P}.
 \een
This enables us to infer that the time evolution of the
corresponding system is pseudo-unitary,
 \ben
 |\psi(t)\rangle = e^{-iHt}\,
 |\psi(0)\rangle
=
\sum_{n=0}^\infty
| \psi_n \rangle \,e^{-i\, E_n\,Q_n\,t} \,
\langle \psi_n |
{\cal P}
 |\psi(0)\rangle ,
 \een
and preserves the value of the scalar product in time,
 \ben
\langle \psi(t)| {\cal P}| \psi(t) \rangle=
\langle \psi(0)| {\cal P}| \psi(0) \rangle.
 \een
We have to stress that the Stone's theorem (which relates the
unitary evolution law to the Hermitian underlying Hamiltonians)
finds the first interesting extension here.

\section{Spontaneously broken ${\cal PT}$ symmetry}

\subsection{Harmonic oscillator inspiration}

Clear parallels between the Hermitian and non-Hermitian ${\cal
PT}$ symmetric Hamiltonians remain marred by the possibility that
in the latter case the reality of the spectrum could break down at
certain couplings. We have seen that ``step-by-step", at an
increasing sequence of the couplings, the levels can cease to be
real even for the most transparent square well and quartic
examples. Here, we are going to explain that from a formal point
of view, the break-down of the ${\cal PT}$ symmetry can still
remain quite innocent in its physical consequences.

When we return to the harmonic oscillator example (\ref{HO}), we
can mimic the break-down of the ${\cal PT}$ symmetry when we
remove the constraint $G > -1/4$ which was inherited from the
standard Hermitian quantum mechanics where its violation would
cause the unavoidable fall of the particles into the attractive
strong singularity in the origin \cite{Landau}.

In the present non-Hermitian context, one cannot find any
persuasive excuse why the smaller couplings $G < -1/4$ could not
be admitted as legitimate. Of course, they give the purely
imaginary parameters $\alpha = i\,\gamma = i\,\sqrt{1/4-G}$ but
the rest of the construction in subsection \ref{HOlab} would
remain perfectly valid. In particular, the termination of the
hypergeometric series will definitely determine the normalizable
solutions existing at the complex energies. This is a puzzle which
is to be resolved here.

Our illustrative harmonic oscillator energies form the two complex
families,
 \be
E_N=4n+2-2\,i\,Q\,\gamma,
 \ \ \ \ \ \ \ \ \ \ \ \gamma = \sqrt{-G-\frac{1}{4}}>0
 , \ \ \ \ \ \ \ \ G < -\frac{1}{4}.
\label{brspectr}
 \ee
These energies (as well as their Laguerre-polynomial wave
functions) are numbered, as above, by the quasi-parity $Q = \pm 1$
and by the integers $n = 0, 1, \ldots$ in the index $N =
2n+(1-Q)/2$.

We are going to demonstrate now that the similar families of the
complex energies can still be interpreted as admissible solutions.
We shall see that, rather counterintuitively, there exists in fact
no acceptable reason why the complex spectrum (\ref{brspectr})
should be forgotten~\cite{Mezincescu}.

We have to return to a model-independent argumentation. In the
case of the broken symmetry, we shall only assume that the two
solutions with $E\neq E^*$ have to be sought simultaneously.

\subsection{The case of the complex conjugate pairs of
the energies}

In a way inspired by ref. \cite{spont} we may assume that the
${\cal PT}$ symmetry of the Hamiltonian becomes broken by a pair
of the wave functions. One gets the two respective Schr\"{o}dinger
equations
 \ben
H |\psi_+\rangle = E\,|\psi_+ \rangle, \ \ \ \ \ \ \ \ \ \ \ H
|\psi_{-}\rangle = E^*\,|\psi_{-} \rangle.
 \een
As long as we have $H \neq H^\dagger={\cal P} \,H \,{\cal P}$, we
may also re-write our equations in the form of action of the
Hamiltonian to the left,
 \ben
\langle \psi_+ | \,{\cal P} \,H  = E^*\, \langle \psi_+ |\,{\cal
P}, \ \ \ \ \ \ \ \ \ \ \ \langle \psi_{-} | \,{\cal P} \,H  = E\,
\langle \psi_{-} |\,{\cal P}.
 \een
Out of all the possible resulting overlaps, let us compare the
following two,
 \ben
\langle
\psi_{+} |\,{\cal P}\,
H |\psi_{+}\rangle = E^*\,\langle
\psi_{+} |\,{\cal P}\,
|\psi_{+} \rangle,
\ \ \ \ \ \ \ \ \ \ \
\langle
\psi_{+} | \,{\cal P} \,H\,|\psi_{+} \rangle  = E\,
\langle
\psi_{+} |\,{\cal P}\,|\psi_{+} \rangle
\een
and, in parallel,
 \ben
\langle \psi_{-} |\,{\cal P}\, H |\psi_{-}\rangle = E^*\,\langle
\psi_{-} |\,{\cal P}\, |\psi_{-} \rangle, \ \ \ \ \ \ \ \ \ \ \
\langle \psi_{-} | \,{\cal P} \,H\,|\psi_{-} \rangle  = E\,
\langle \psi_{-} |\,{\cal P}\,|\psi_{-} \rangle.
 \een
These alternatives imply that for $E\neq E^*$ the self-overlaps
must vanish,
 \ben
\langle \psi_+ |\,{\cal P}\, |\psi_+\rangle = 0, \ \ \ \ \ \ \ \ \
\ \ \langle \psi_{-} |\,{\cal P}\, |\psi_{-}\rangle = 0.
 \een
This leads to several interesting consequences.  Firstly, we are
free to employ the following less common normalization
 \ben
\langle
\psi_+ |\,{\cal P}\, |\psi_{-}\rangle =
\left [ \ \langle
\psi_{-} |\,{\cal P}\, |\psi_{-}\rangle \ \right ]^*= c,
 \een
and, wherever needed, re-normalize $c \to \pm 1$.  This convention
is less common but can be still interpreted as a generalized
orthonormality condition in any two-dimensional subspace of the
linear pseudo-normalized space of the ${\cal PT}$ symmetry
breaking states.

\subsection{Evolution under the broken ${\cal PT}$
symmetry}

For the sake of simplicity, let us assume that the ${\cal PT}$
symmetry is broken just at the two lowest states.  The necessary
modification of the completeness relations adds then just the
two new terms to the sum over the unbroken $\psi_n$'s,
 \ben
I = | \psi_+ \rangle \, \frac{1}{c^*}\, \langle \psi_{-} | {\cal
P}\  + \ | \psi_{-} \rangle \, \frac{1}{c}\, \langle \psi_+ |
{\cal P}\ + \ \sum_{n=2}^\infty | \psi_n \rangle \, Q_n \langle
\psi_n | {\cal P}  .
 \een
The forthcoming modification of the spectral decomposition of the
Hamiltonian adds the similar two new terms to the sum over the
unbroken energies,
 \ben
I = | \psi_+ \rangle \, \frac{E}{c^*}\, \langle \psi_{-} | {\cal
P}\  + \ | \psi_{-} \rangle \, \frac{E^*}{c}\, \langle \psi_+ |
{\cal P}\ +\ \sum_{n=2}^\infty | \psi_n \rangle \,E_n Q_n \langle
\psi_n | {\cal P}  .
 \een
Finally, the pseudo-unitary time development acquires the compact
form as well,
 \ben
 |\psi(t)\rangle = e^{-iHt}\,
 |\psi(0)\rangle
=\een
 \ben
 | \psi_+ \rangle \, \frac{1}{c^*}\, e^{-i\,E\,t}\, \langle
\psi_{-} | {\cal P}\  + \ | \psi_{-} \rangle \, \frac{1}{c}\,
e^{-i\,E^*\,t}\, \langle \psi_+ | {\cal P} \ +\ \sum_{n=2}^\infty
| \psi_n \rangle \,e^{-i\, E_n\,Q_n\,t} \, \langle \psi_n | {\cal
P}
 |\psi(0)\rangle
 .
 \een
It is quite amusing to discover that the value of the scalar
product is conserved,
 \ben
\langle \psi(t)| {\cal P}| \psi(t) \rangle=
\langle \psi(0)| {\cal P}| \psi(0) \rangle.
 \een
A full parallel to the conventional quantum mechanics is
established.

\section{Summary}

We may summarize that far beyond the boundaries of the ordinary
quantum mechanics, the above-mentioned difficulties with the
implications of the Stone's theorem in ${\cal PT}$ symmetric
context were shown related to the ``hidden" use of a pseudo-norm.
{\it Vice versa,} after one admits that the vanishing
(pseudo-)norm need not imply the vanishing of the state, a
modified version of the Stone's theorem is recovered.  We have
shown that the appropriately defined self-overlaps of the states
remain unchanged in time not only in the systems characterized by
the preserved ${\cal PT}$ symmetry, but also in the domains of
couplings where this symmetry is spontaneously broken.

A consistent and complete interpretation of the extended quantum
formalism is not at our disposal yet.  Still, several new features
of it have been revealed here.  For example, it seems to mimic
some properties of the indefinite metric which is already of quite
a common use, say, in relativistic physics.

\section*{Acknowledgement}

An e-mail-mediated discussion with I. Herbst and G. A. Mezincescu
proved inspiring.  Partially supported by the GA AS grant Nr. A
104 8004.


\newpage
\begin{thebibliography}{99}

\bibitem{Landau}
L. D. Landau and E. M. Lifshitz, Quantum Mechanics (Pergamon,
London, 1960).

\bibitem{BBjmp}
C. M. Bender, S. Boettcher and P. N. Meisinger, J. Math. Phys. 40
(1999) 2201;

C. M. Bender, S. Boettcher and M. Van Savage, J. Math. Phys. 41
(2000) 6381.

\bibitem{BMa}
C. M. Bender and K. A. Milton, Phys. Rev. D 57 (1998) 3595.

\bibitem{SQW}
M. Znojil, ``PT symmetric square well" (arXiv: quant-ph/0101131),
submitted.

\bibitem{hopt}
M. Znojil, Phys. Lett. { A 259} (1999) 220.

\bibitem{Caliceti}
E. Caliceti, S. Graffi and M. Maioli, Commun. Math. Phys. 75
(1980) 51.

\bibitem{Alvarez}
G. Alvarez, J. Phys. A: Math. Gen. 27 (1995) 4589;

M. Znojil, J. Phys. A: Math. Gen. 32 (1999) 7419;

B. Bagchi and R. Roychoudhury, J. Phys. A: Math. Gen. 33 (2000)
L1;

M. Znojil and G. L\'{e}vai, Phys. Lett. { A 271} (2000) 327;

E. Delabaere and D. T. Trinh, J. Phys. A: Math. Gen. 33 (2000)
8771.

\bibitem{ixna}
F. Fern\'andez, R. Guardiola,  J. Ros and M. Znojil,  J. Phys. A:
Math. Gen. 31 (1998) 10105.

\bibitem{BG}
V. Buslaev and V. Grecchi, J. Phys. A: Math. Gen. 26 (1993) 5541.

\bibitem{BT}
C. M. Bender and A. Turbiner, Phys. Lett. A 173 (1993) 442;

A. A. Andrianov, F. Cannata, J. P. Dedonder and M. V. Ioffe, Int.
J. Mod. Phys. A 14 (1999) 2675;

M. Znojil, F. Cannata, B. Bagchi and R. Roychoudhury, Phys. Lett.
B 483 (2000) 284;

B. Bagchi, S. Mallik and C. Quesne, Int. J. Mod. Phys. A, to
appear (arXiv: quant-ph/0102093).

\bibitem{QED}
C. M. Bender and K. A. Milton,  J. Phys.  A: Math. Gen. 32 (1999)
L87.

\bibitem{Pham}
E. Delabaere and F. Pham, Phys. Letters A 250 (1998) 25 and 29;

P. Dorey, C. Dunning and R. Tateo, ``Spectral equivalences from
Bethe Ansatz equations" (arXiv: hep-th/0103051), submitted.

\bibitem{BB}
C. M. Bender and S. Boettcher, Phys. Rev. Lett. { 24} (1998) 5243.

\bibitem{QES}
C. M. Bender and S. Boettcher, J. Phys.  A: Math. Gen. 31 (1998)
L273;

M. Znojil, J. Phys. A: Math. Gen. 32 (1999) 4563;

B. Bagchi, F. Cannata and C. Quesne, Phys. Lett. A 269 (2000) 79;

M. Znojil, J. Phys. A: Math. Gen. 33 (2000) 4203 and 6825;

F. Cannata, M. Ioffe, R. Roychoudhury and P. Roy, ``A new class of
PT-symmertric Hamiltonians" (arXiv: quant-ph/0011089), submitted.

\bibitem{exact}
F. Cannata, G. Junker and J. Trost, Phys. Lett. { A 246} (1998)
 219;

M. Znojil, Phys. Lett. { A 264} (1999) 108.

B. Bagchi and C. Quesne, Phys. Lett. A 273 (2000) 7165;

M. Znojil, J. Phys. A: Math. Gen. 33 (2000) L61 and 4561;

G. L\'{e}vai and M. Znojil, J. Phys. A: Math. Gen. 33 (2000) 7165.

\bibitem{Bessis}
Daniel Bessis, private communication;

M. Znojil and M. Tater, J. Phys. A: Math. Gen. 34 (2001) 1793.

\bibitem{BMb}
C. M. Bender and K. A. Milton, Phys. Rev. D 55 (1997) R3255.

\bibitem{Simsek}
Bender C M, Berry M, Meisinger P N, Savage V M and Simsek M 2001
J. Phys. A: Math. Gen. 34 L31.

\bibitem{creation}
M. Znojil, ``Annihilation and creation operators anew" (arXiv:
hep-th/0012002), submitted.

\bibitem{whatis}
M. Znojil, ``What is PT symmetry?" (arXiv: quant-ph/0103054),
submitted.

\bibitem{Mezincescu}
G. A. Mezincescu,  J. Phys. A: Math. Gen. 33 (2000) 4911 and
private communication.

\bibitem{spont}
A. Khare and B. P. Mandel, Phys. Lett. A 272 (2000) 53.

\bibitem{Seiler}
R. Seznec and J. Zinn-Justin, J. Math. Phys. 20 (1979) 1398;

J. Avron and R. Seiler, Phys. Rev. D 23 (1981) 1316;

A. Andrianov, Ann. Phys. 140 (1982) 82.

\bibitem{Dyson}
F. J. Dyson, Phys. Rev. 85 (1952) 631.

\bibitem{Herbst}
I. Herbst, private communication.

\bibitem{Fluegge}
S. Fl\"{u}gge, Practical Quantum Mechanics I (Springer, Berlin,
1971), p. 166.

\end{thebibliography}
\end{document}